\documentclass{article}

\usepackage{PRIMEarxiv}
\usepackage{longtable}
\usepackage[utf8]{inputenc} 
\usepackage[T1]{fontenc}    
\usepackage{hyperref}       
\usepackage{url}            
\usepackage{booktabs}       
\usepackage{amsfonts}       
\usepackage{nicefrac}       
\usepackage{microtype}      
\usepackage{lipsum}
\usepackage{fancyhdr}       
\usepackage{graphicx}       
\graphicspath{{figure/}}     

\title{MobileAgent: Enhancing Mobile Control via Human-Machine Interaction and SOP Integration
}

\author{
  Tinghe Ding\textsuperscript{\rm}, 
    Yang Wu\textsuperscript{\rm}, 
    Chenyi Zhuang\textsuperscript{\rm}, 
    Xiaodong Zheng\textsuperscript{\rm},
    Guannan Zhang\textsuperscript{\rm}\\
    Ant Group\\
    \texttt\{tinghe.dth, wy306396, chenyi.zcy, xiaodong.zxd, zgn138592\}@antgroup.com
}

\begin{document}
\maketitle

\begin{abstract}

Agents centered around Large Language Models (LLMs) are now capable of automating mobile device operations for users. After fine-tuning to learn a user's mobile operations, these agents can adhere to high-level user instructions online. They execute tasks such as goal decomposition, sequencing of sub-goals, and interactive environmental exploration, until the final objective is achieved. However, privacy concerns related to personalized user data arise during mobile operations, requiring user confirmation. Moreover, users' real-world operations are exploratory, with action data being complex and redundant, posing challenges for agent learning. To address these issues, in our practical application, we have designed interactive tasks between agents and humans to identify sensitive information and align with personalized user needs. Additionally, we integrated Standard Operating Procedure (SOP) information within the model's in-context learning to enhance the agent's comprehension of complex task execution. Our approach is evaluated on the new device control benchmark AitW, which encompasses 30K unique instructions across multi-step tasks, including application operation, web searching, and web shopping. Experimental results show that the SOP-based agent achieves state-of-the-art performance in LLMs without incurring additional inference costs, boasting an overall action success rate of 66.92\%. The code and data examples are available at https://github.com/alipay/mobile-agent.
\end{abstract}


\section{Introduction}
\begin{figure*}[t]

    \vspace{0.2cm}
    \centering
    \includegraphics[width=\textwidth,keepaspectratio]{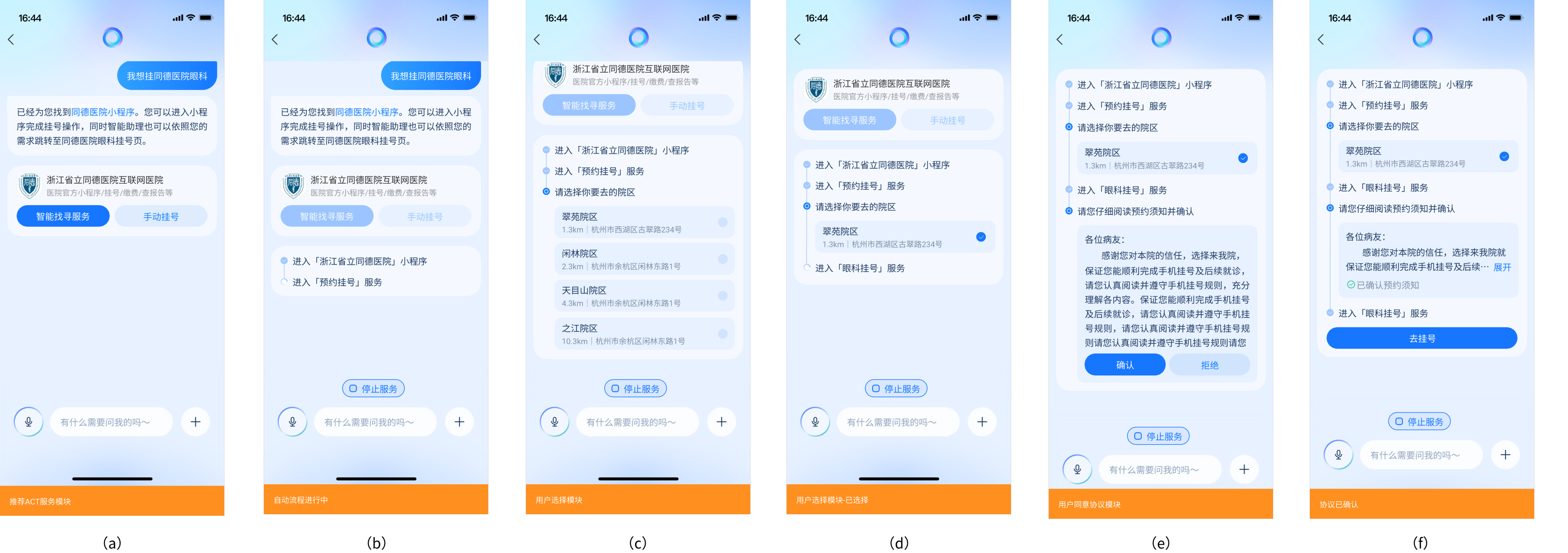} 
    \caption{The overview describes an automated operational tool app used in a chat scenario for scheduling hospital appointments. The process involves:
    (a) The user commanding the AI agent in the chat to schedule an appointment.
    (b) The AI agent navigating the hospital app to book an appointment.
    (c) The agent identifying different hospital branches, organizing this information, and displaying it in the chat.
    (d) The user selecting a branch in the chat, and the agent interacting with the app accordingly.
    (e) The agent recognizing and presenting essential pre-appointment details in the chat.
    (f) The user confirms, and the agent completes subsequent steps.
    }
    \label{myframework}
\end{figure*}
Due to the advanced capabilities and increasing popularity of LLMs \cite{ouyang2022training,wei2022emergent}, researchers are now using these models to develop AI agents with enhanced perceptual and action abilities, employing strategies such as multimodal perception and tool utilization \cite{park2023generative,deng2023mind2web,burns2022dataset,sumers2023cognitive,yao2022webshop,wang2023survey,xi2023rise}. These agents facilitate user interactions with electronic devices through clicks, gestures, and text, thereby streamlining complex tasks \cite{rawles2023android,wen2023empowering,zhan2023you,yan2023gpt}.
Significant research in this domain includes web-based projects like WebGum\cite{furuta2023multimodal}, Webagent\cite{gur2023real}, Mind2Web\cite{deng2023mind2web}, Webshop\cite{yao2022webshop}, and mobile-focused initiatives such as Auto-UI\cite{zhan2023you}, BC agent \cite{rawles2023android}.
The development of these intelligent agents typically involves task planning, environment perception, intent recognition, and understanding operation sequences to make execution decisions.
These decisions encompass determining task statuses and generating operation types\cite{rawles2023android}.
Current research highlights three critical factors influencing their practical application on user mobile.
\begin{itemize}
    \item Firstly, it is challenging for agents to establish a relationship between current page operations and task objectives from fragmented and redundant user operational behaviors. User operational behaviors available for agent learning are characterized by a degree of arbitrariness, necessitating that the agent comprehends historical user behavior to refine its automated execution skills \cite{rawles2023android,zhan2023you}. Training samples often consist of complex or repetitive user actions. For example, a user's persistent scrolling and browsing during a purchase, or their repeated searches and browsing for related queries, are typical of such behaviors. To efficiently process these intricate actions, the large model must accurately identify the essential task pipeline throughout the entire task execution.
    \item Moreover, some privacy-related decisions cannot be fully automated, as they necessitate user awareness and confirmation. As agent capabilities advance, human involvement becomes increasingly crucial to guide and monitor agents’ actions, ensuring they align with human requirements and objectives \cite{wang2023survey,xi2023rise,shuster2022blenderbot,henderson2018ethical}. Often, in real-world scenarios, machines do not need to perform certain tasks autonomously, particularly when handling sensitive information. As depicted in Figure \ref{myframework}, our developed application automates operations based on user commands in a chat scenario. For instance, scheduling a hospital appointment via an app must involve user input for privacy authorization, clinic selection, and receiving pre-visit guidelines.
    \item The ultimate challenge in enhancing predictive methods often leads to increased computational demands during online deployment. For example, integrating smaller models to complement the LLM with additional information, or employing techniques such as planning and chaining actions for predictions, can significantly slow down the prediction process \cite{gur2023real,song2023llm,zhang2023igniting}. This slowdown is contrary to the essential principle of rapid execution, which is crucial for large-scale automation models.
\end{itemize}
In response, we have developed a more versatile range of execution types for practical applications. This involves identifying key information and structuring it for user interaction, aligning more closely with real-world scenarios. 
For more advanced task instructions, the agent must understand the intent and then decompose it into a series of subtasks, where their understanding of the tools significantly influences the decomposition process. Furthermore, SOP \cite{wagner2014sop,tummala2004sop,hong2023metagpt,zhou2023agents}, defined as detailed instructions for consistency and quality in complex tasks, are pivotal in automation. We have integrated a vital operation pipeline and the monitoring of sub-task execution statuses within the current environment. This approach has improved predictive performance without increasing online prediction costs.

\section{Related Work}

AI agent, leveraging a substantial language model as its core processing unit, is capable of engaging in dialogues, executing tasks, reasoning, and displaying a degree of independent action. To augment its autonomy, such an agent requires integration with extensive databases, a memory system, and access to reasoning tools. Prompt engineering equips these agents with advanced capabilities for analysis, planning, implementation, evaluation, and continuous improvement. With appropriate information and prompts, they can manage self-sufficient workflows, albeit under some human supervision\cite{wang2023survey,xi2023rise}.\\
\begin{figure}[h]
    \centering
    \includegraphics[width=0.47\textwidth]{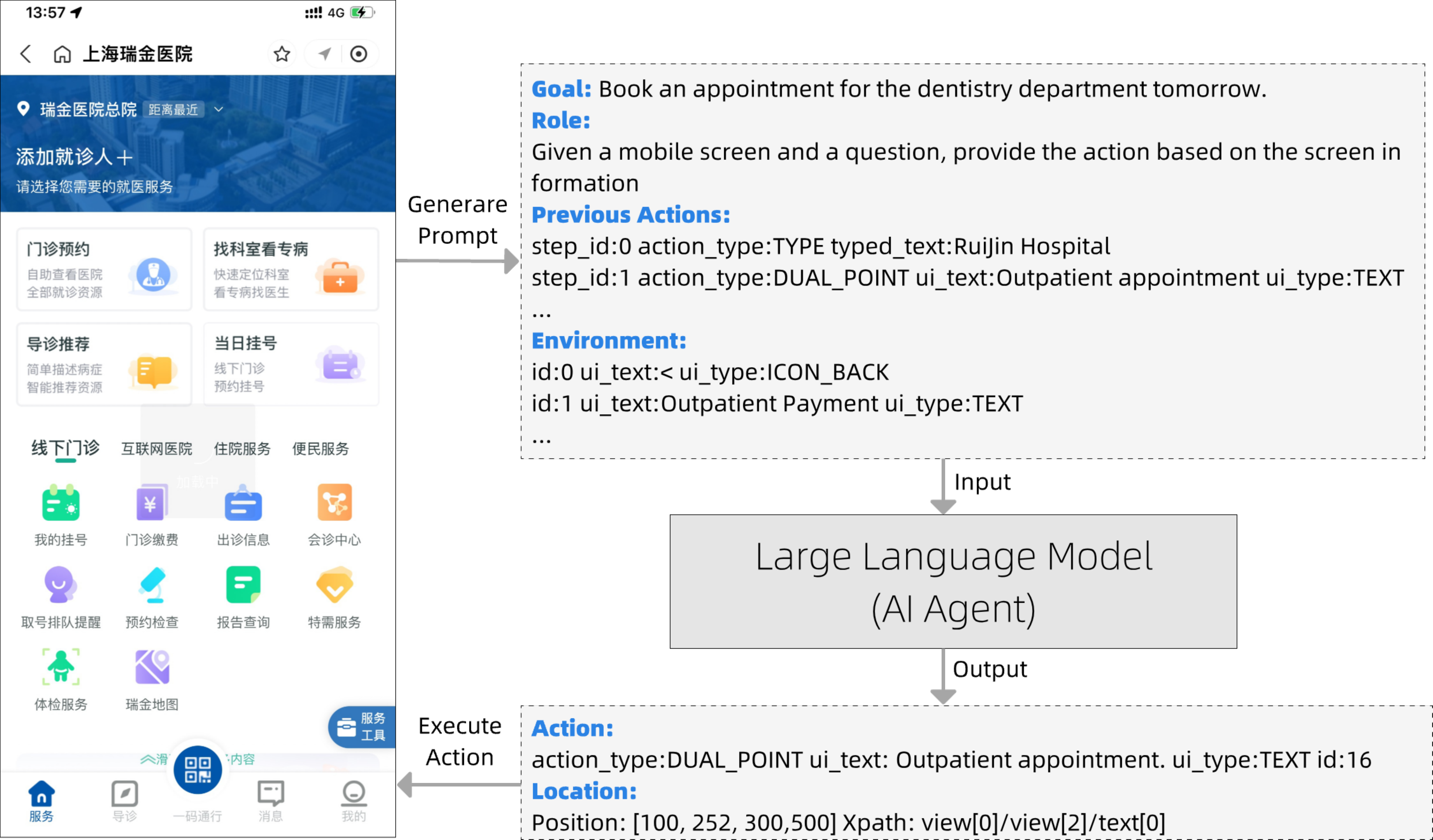} 
    \caption{Overview of the process of an automated execution tool. Extracting information from a mobile page as DOM context.
    Subsequently, the task goal, agent role, DOM information, and historical operations are combined to form the model's input prompt.
    Inputting the prompt to the AI agent to generate execution commands, and carrying out operations on the mobile page based on these commands.}
    \label{mypipeline}
\end{figure}
Regarding manipulation tasks for agent, there are two particularly popular areas: web page manipulation and mobile application operation\cite{park2023generative,deng2023mind2web,burns2022dataset,sumers2023cognitive,yao2022webshop}. 
Overall, the task-oriented agent follows high-level user instructions, performing tasks like goal decomposition \cite{gur2023real,wu2023plan}, planning the sequence of sub-goals\cite{zhan2023you,wu2023plan,chen2023interact}, and interactive exploration of the environment \cite{yao2022webshop,furuta2023multimodal,zhou2023webarena}, until the final objective is attained. Figure \ref{mypipeline} is an example of a basic operation instruction generation for mobile application operation.\\
Agents must have the capability to comprehend instructions in complex scenarios, adapt to changes like noisy text and dynamic HTML web pages or DOM application pages, and gradually generalize successful operations to ensure the completion of advanced task objectives through extended operational processes\cite{zhou2023webarena}.
Researchers have begun leveraging the advanced HTML reading and visual perception abilities of LLM, such as understanding entire page source codes \cite{deng2023mind2web} and employing a multimodal corpus containing screenshots \cite{zhan2023you,furuta2023multimodal,hong2023cogagent,yang2023appagent}, to enhance understanding capabilities.\\
In addition to generating the final operational objective, specific details such as element coordinates or XPath (XML Path Language) \cite{clark1999xml} are crucial for task execution. Various models adopt different approaches: some integrate these coordinates as both input and output \cite{rawles2023android,zhan2023you}, while others focus on predicting the element objects and employ strategies to align them with the corresponding coordinates \cite{deng2023mind2web,yan2023gpt}.\\
In the process of observing the execution environment, maintaining synchronization between the agent's state and the subgoals is significant.
Some researchers employ agent to decompose high-level task instructions into a series of sub-goals \cite{chen2023interact}, basic skill sequences \cite{nottingham2023embodied}, or fundamental keyboard/mouse operations \cite{yuan2023plan4mc}. 
In terms of control strategies, planning the sequence of sub-goals requires decomposing complex tasks into multiple subtasks. This not only enables a better understanding of the intent behind user operations but also allows for the accomplishment of a significantly greater workload \cite{wang2023survey,xi2023rise}.\\
When handling tasks related to sensitive information, involving agents in collaboration with humans is an effective approach. Humans can offer guidance and help regulate the safety, legality, and ethical conduct of agents \cite{xi2023rise}. 
For example, in travel decision-making, agents can enhance the experience by empowering tourists to actively seek highly relevant information through a question-and-answer mode, essentially placing them in the driver's seat of the decision process \cite{wong2023autonomous}.
But it is challenging to ascertain privacy information during an agent's execution and to proactively inform users through dialogue. Consequently, agents must be capable of dynamically recognizing emerging privacy information within the operational environment. Subsequently, this information should be structured and presented to users for confirmation of the execution details.

\section{Method}

In this field, use LLM to plan the sequence of sub-goals increases the perplexity of inference. For instance, in our collected training tasks, the same task and environment can have multiple execution pipelines. In such tasks, LLMs struggle to generate consistent sub-goal sequences. If recall techniques are used to place related subtask pipelines into the LLM's in-context, this approach faces issues with recall accuracy and also increases the time consumed in the execution chain.
Therefore, we will discuss how introducing SOP method can enhance model Performance. Additionally, We developed structured instructions to identify and extract privacy information, thereby increasing human interaction, and we implemented the latest data processing methods based 'Enlarged-element' operations.
We will address this topic focusing on Sample Handling, Structured Instruction, and SOP Design.\\
\textbf{Sample Handling:} Many existing methods incorporate the coordinates of page elements into the model's input, whereas the coordinates of user clicks or the start and end points of swipes are processed as outputs. Despite the numerical connection between these two types of coordinates, accurately discerning their relationships poses a challenge for large models. To address this, we employ 'Enlarged-element' operations, which map coordinates to specific page elements by expanding each element's coordinate bounding box. This approach ensures unique association of click coordinates with designated elements. In the case of click operations, the coordinates are mapped to specific elements, which are then identified as the predicted targets. For swipe operations, we limit the output to the swipe direction, with each direction corresponding to a predetermined sliding distance. The predictive targets for each operation type and their corresponding coordinate transformations are detailed in Appendix Table \ref{target_output}.\\
\textbf{Structured Instruction:} To ensure user privacy and address personalized needs, we have developed structured instructions in addition to the common operational and task status instructions. For application pages that require user awareness and authorization, the agent is designed to accurately identify and enable the production of well-structured messages and user operation choices. This empowers users to select operations themselves, effectively replacing the model's direct execution of commands. Examples of all types of instructions can be found in the Appendix Figure \ref{instructions}.\\
\textbf{SOP Design:} MetaGPT \cite{hong2023metagpt} introduces a groundbreaking approach by leveraging SOP to orchestrate multi-agent systems driven by LLM, revolutionizing collaborative task resolution. In our research, we utilize SOPs to break down complex tasks into more manageable segments, assigning subtasks and monitoring their completion status within the current action environment. Unlike traditional plan-based methods, SOPs function as a blueprint for the model’s instruction execution pipeline on the input side, while a plan dictates future action strategies during the output phase. SOPs abstract and generalize operational instructions, eschewing the specification of explicit operation types.
To execute a task such as 'Search for the best rated headphones on Amazon,' the SOP can be structured as follows: First, conduct a search on the website; second, view and click on page content to access Amazon; third, type 'Best Rated Headphones' into Amazon's search bar; fourth, view and select relevant products by clicking on their page content; finally, complete the task.
We also integrate the current status of task execution into the SOPs, enabling the model to more effectively understand past actions and anticipate future tasks.\\
In our approach, we introduce an abstraction \( Y \) that encapsulates various types of instructions generated from the model's input prompt \( X \). The input prompt includes the task goal, agent role, DOM information, and historical operations. Mathematically, we can formalize this process as follows:
The conditional entropy of directly predicting \( Y \) given \( X \) is denoted as \( H(Y|X) \). Upon introducing the abstract SOP representation \( Z \), we define the conditional entropy for predicting \( Y \) with the knowledge of both \( X \) and \( Z \) as \( H(Y|X, Z) \).
According to the chain rule of information theory, it is known that:
\[ H(Y|X, Z) \leq H(Y|X) \]
This implies that the uncertainty of \( Y \), given both \( X \) and \( Z \), is potentially reduced compared to when only \( X \) is known. Therefore, the presence of \( Z \) can lead to a reduction in entropy—or equivalently, an increase in the information available for the prediction of \( Y \)—which can consequently improve the accuracy of the prediction.

\section{Experiments}
We have enhanced our Ant Intelligent Assistant (AIA) application by integrating the capability of large models to recognize privacy permissions and generate structured information, thereby improving the functionality of user interactions in automation execution. Furthermore, we have developed a LLM agent based on SOP technology, utilizing Google's AitW dataset \cite{rawles2023android}, to objectively assess the accuracy and time efficiency of our method.

\subsection{AIA}
The AIA is a product currently in the testing and development phase, not yet released for public use, designed and developed for dialogue scenarios where large models generate instructions to aid user automation in mobile applications. Within AIA, we have formulated Operational Instructions, Task Status Instructions, and Structured Instructions to enhance user interaction, the interactions example can be found in Appendix Figure \ref{instructions}.
From the model’s perspective, the input is a prompt comprising task goals, operation sequences, and the DOM information of the current page. The model outputs three distinct types of instructions.
\begin{itemize}
\item     Operational Instructions: These include commands such as clicking, scrolling, and typing, which facilitate basic mobile phone operations.

\item     Task Status Instructions: These instructions represent the completion of a task, indicating the status of the instruction, such as successful completion or infeasibility.
    
\item     Structured Instructions: This category involves identifying necessary user interactions on the current page and transforming disorganized DOM fragment data into a well-structured data format.    
\end{itemize}

\begin{figure}[]
    \centering
    \includegraphics[width=0.47\textwidth]{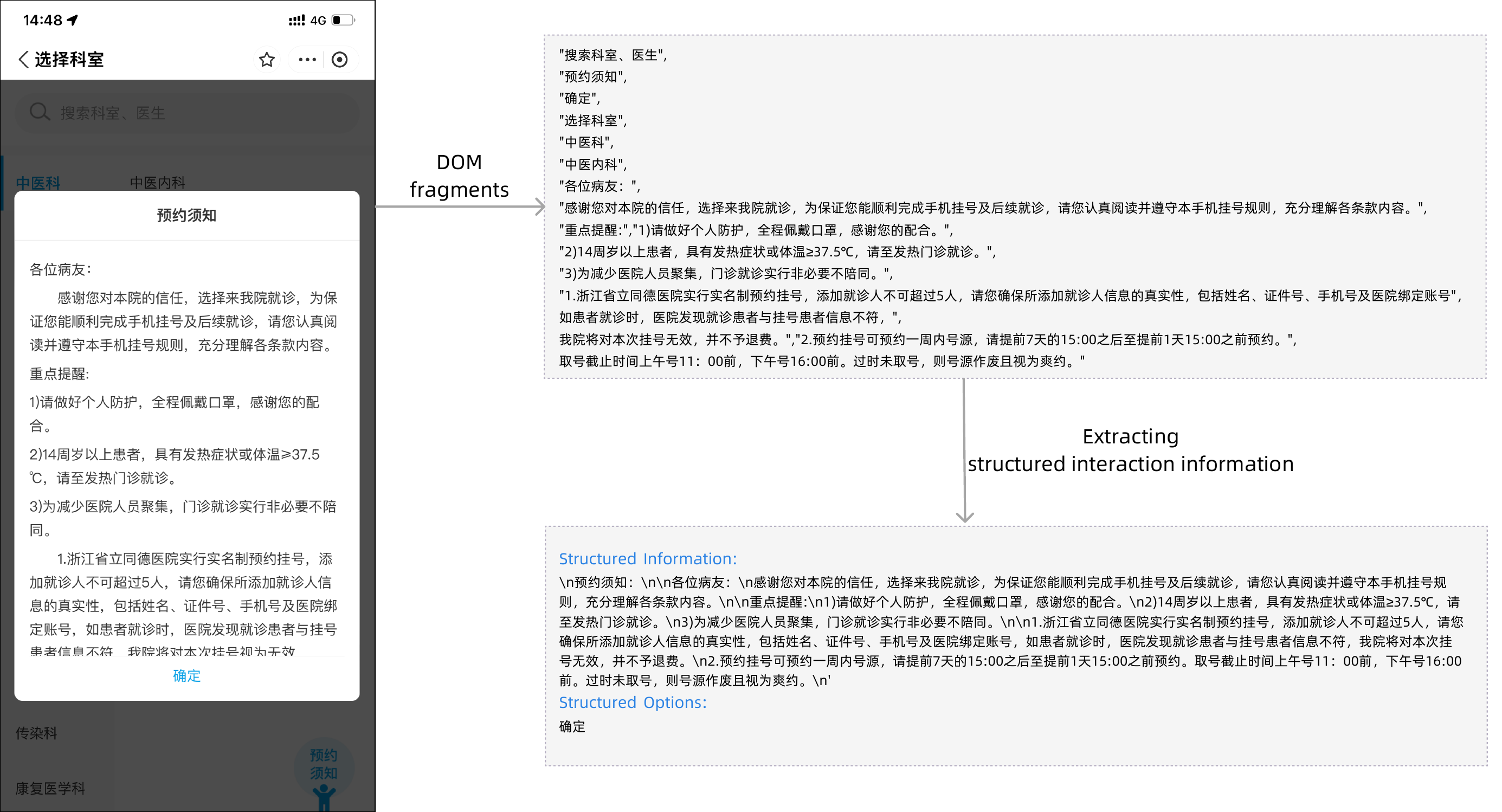} 
    \caption{The prediction process in the notification confirmation scenario.}
    \label{structed}
\end{figure}

\paragraph{Strctured Instructions}
Exploring the Structured Instruction functionality, this aspect focuses on identifying pages that necessitate the recognition of personal privacy information authorization, the confirmation of message notifications, and the selection of slot content. As illustrated in Figure \ref{structed}, when processing a medical page with pop-up information, including medical instructions, we start by organizing the fragmented and disorganized DOM data. This cleaned-up data is then used to formulate a prompt for the AI agent. Subsequently, the AI agent generates structured notification content and user-required confirmation options, such as 'I understand'.

\paragraph{SOP Details}
In our AIA medical application, the primary action involves clicking on text elements. We have delineated 13 fundamental SOP sub-tasks within the medical in-context, including Select Department, Others, Consent Information, Select Clinic Area, Order Payment, Select Appointment Time, Outpatient Registration, Outpatient Payment, View Reports, Check Expenses, Registration History, Electronic Invoices, and Inpatient Services, each linked to specific text elements for interaction. Utilizing this framework, we trained a BERT-based classification model \cite{devlin2018bert}. In this setup, each click text in the user's action sequence is categorized into a corresponding sub-task, culminating in the formation of the SOP pipeline. For instance, phrases like 'confirm', 'I understand', 'agree to authorize', and 'log in' are classified by the BERT model into the 'Consent Information' sub-task, while text elements such as 'urology', 'general surgery', and 'reproductive clinic' fall under the 'Select Department' sub-task. Thus, every click text is assigned to a pertinent SOP task, constructing an SOP pipeline. However, click texts categorized under the 'Others' task are excluded from the SOP pipeline.
\paragraph{Evaluation Measures}
Task completion rate is utilized as a metric to evaluate the effectiveness of predictions. Initially, a partial score is determined by dividing the number of correct actions by the episode length. Subsequently, the task completion score is computed as the average of all partial scores.
\paragraph{Results}
In response to privacy concerns, the use of online user data was ruled out. Consequently, we engaged ten external contractors to conduct testing operations over a two-week period. Some of these participants were assigned the responsibility of manually documenting their envisioned utilization of medical scenarios within the Alipay app, while others operated the Alipay application to simulate these specific scenarios. Following this, we collected their data with the aim of constructing model samples. Additionally, we collected corresponding tasks from 38 hospital applications, amassing a total of approximately 20,000 episodes. Of these, around 2,000 episodes were allocated for testing purposes. Regarding the model, we utilized QWen-7B \cite{bai2023qwen} to conduct Supervised Fine-Tuning (SFT) with LoRA on our chinese dataset. The related evaluation metrics are detailed below in Table \ref{act_result}. Integrating SOP content within the model's in-context learning significantly improves the accuracy of instruction generation.

\begin{table*}[h]
    \centering
\begin{tabular}{lll}
  \toprule
  Instruction & Qwen-7B & Qwen-7B-SOP  \\
  \midrule
  Operational    & 0.895  & \textbf{0.926} \\
  Task Status  & 0.792   & \textbf{0.942}  \\
  Structured   & 0.46 & \textbf{0.606}  \\
  \bottomrule
\end{tabular}
\caption{Diverse Instruction Types and Their Corresponding Impact on Models' Predictive Performance (task complete score).}
\label{act_result}
\end{table*}

\subsection{AitW}
The AitW dataset \cite{rawles2023android} is highly regarded in the domain of mobile device control, comprising human-curated demonstrations of natural language instructions, user interface (UI) screens, and actions across a range of human tasks. AitW is a well-recognized dataset in the field of mobile operations, extensively utilized by numerous academic institutions and research organizations for research purposes. The AitW task example show as Appendix Figure \ref{example_aitw}. To objectively assess the effectiveness of our SOP technology in enabling large models to quickly learn tool usage, we conducted evaluations using this dataset.
\paragraph{Dataset}
The dataset comprises human demonstrations of device interactions, encompassing screens, actions, and corresponding natural language instructions. It contains 715,000 episodes, covering 30,000 unique instructions. The benchmark dataset is segmented into five subsets: General, Install, GoogleApps, Single, and WebShopping. Each subset is further divided into training, validation, and test sets on an episode-wise basis. Detailed descriptions and statistics for the subsets of the benchmark dataset are provided in Table \ref{aitw_statics}.
\begin{itemize}
    \item     GOOGLEAPPS: This subset encompasses high-level tasks, some of which overlap with PixelHelp, and involves interactions with various Google applications like Gmail, Calendar, Photos, Settings, etc.

    \item     INSTALL: This category includes tasks related to installing and uninstalling apps, managing app logins, and addressing support issues (e.g., "forgot password") for 88 applications on the Google Play Store.
    
    \item     WEBSHOPPING: Focused on E-commerce activities, this subset covers tasks such as searching for products, adding items to a shopping cart, and reviewing cart contents.
    
    \item     GENERAL: A diverse collection of tasks, primarily centered on question-answering (e.g., "How much does a 2-bedroom apartment rent for in San Francisco?") and interactions with third-party apps and websites.
    
    \item     SINGLE: Comprised of single-step tasks, annotated with hindsight relabeling, predominantly sourced from the WEBSHOPPING category (e.g., "Close the pop-up, then add the first item to the cart").
\end{itemize}

\begin{table*}[]
    \centering
\begin{tabular}{lllll}
    \toprule
    Dataset & Episodes & Screens & Instructions \\
    \midrule
    General & 9,476  & 85,413 & 545 \\
        Install & 25,760 & 250,058 & 688  \\
        GoogleApps & 625,542 & 4,903,601 & 306  \\
        Single & 26,303 & 85,668 & 15,366  \\
        WebShopping & 28,061 & 365,253 & 13,473  \\
        \bottomrule
    \end{tabular}
    \caption{Dataset statistics.}
    \label{aitw_statics}
\end{table*}

\paragraph{Baseline}
We employed various baseline models, specifically:
\begin{itemize}
  \item  In-context learning LLMs: This includes Palm 2\cite{anil2023palm} and ChatGPT-3.5\cite{abdullah2022chatgpt}. These models were used without a fine-tuning process but with extra samples. They input a textual description of the screen in HTML syntax, incorporating UI element information from OCR and icon detection. The models are tasked with predicting actions from predefined options, such as providing the index of the clicked UI element for click actions or specifying the scroll direction for scrolling actions. In the ChatGPT-3.5 model, 5-shot CoT prompting is utilized to enhance performance.

  \item    Fine-tuned LLMs: For this category, we selected Llama 2\cite{touvron2023llama} as the baseline and fine-tuned it using LoRA. The model receives user instructions and screen descriptions, similar to the format used for in-context learning LLMs. The expected output from the model includes the action type and the specific element. For details on the sample design, please refer to Figure \ref{mypipeline} above. Training methods similar to SOP were also employed for control comparisons. This included incorporating the execution plan as well as integrating both the plan and task status into the output, we name them Llama 2+plan and Llama 2+plan+state.
    
  \item    Multimodal LLMs: This category includes ChatGPT-4V\cite{yang2023dawn}. For ChatGPT-4V, screen images are directly utilized, exploring the integration of historical behaviors or OCR-decoded text.
    
  \item    SOP-based LLMs: Similar to the fine-tuned LLMs, we also used Llama 2, based on enlarged-element operations, as the foundational model. In the training phase of the SOP model, we incorporated the task's SOP process into the in-context input and mixed it with the original samples. During the prediction phase, this approach allows consistency with the original prompt, eliminating the need for additional tasks. This mixed training method does not increase the length of the output tokens, ensuring that the time taken for prediction remains unchanged.
\end{itemize}

\paragraph{LLMS Details}
Figure \ref{prompt_statics} clearly illustrates the differences between various techniques by presenting the model's input and output.
\begin{itemize}
\item The topmost Llama 2 model serves as the basic baseline version. Its input prompt consists of task role, previous action, Environment (structured DOM info), and instruction. The output response includes action type and element type (text, type).
\item The Llama 2+plan model adds the complete plan pipeline for executing the instruction to the output response. In this in-context, the term 'task plan' refers to the abstract concept of subtasks, rather than the detailed step-by-step operations.
\item The Llama 2+plan+state model adds information about whether the plan has been completed to the output plan.
\item The Llama 2+SOP approach involves adding key abstract subtask pipelines to the input prompt while also marking the completion status of each subtask in the current process. The method of incorporating SOP in the input in-context is quite similar to adding plan and state in the output.
\end{itemize}

\begin{table*}[h]
    \begin{tabular}{p{1.5cm}p{8.7cm}p{7.1cm}}
    \toprule
    Model & Prompt & Response \\
    \toprule
     {Llama 2} & Given a mobile screen and a question, provide the action based on the screen information. & {action: TYPE} \\
    & Previous Actions:id:0,type:DUAL\_POINT,text:G, ui\_type:... & \\
    & Environment: id:0, text:Bass Headsets, type:TEXT ...& text: best rated headphones\\
    & Instruction: Search for the best rated headphones on Amazon. & \\
    & Answer: & \\
    \midrule
    {+plan} &Given a mobile screen and a question, provide the action based  & {\textbf{PLAN:}} \\
    Llama 2& on the screen information.  & \textbf{id:0 search on the website}  \\
    & Previous Actions:id:0,type:DUAL\_POINT,text:G, ui\_type: ...& \textbf{id:1 view and click page content} \\
    & Environment: id:0, text:Bass Headsets, type:TEXT ... & \textbf{id:2 type 'best rated headphones'} \\
    & Instruction: Search for the best rated headphones on Amazon. & \textbf{id:3 view and click page content}\\
    & Answer: & \textbf{id:4 task complete}\\
    &   & action: TYPE \\
    &   & text: best rated headphones \\
    
    \midrule
    
    {+plan+state} &Given a mobile screen and a question, provide the action based  & {\textbf{PLAN\&STATE:}} \\
    Llama 2& on the screen information.  & \textbf{id:0 search on the website,state:finish}  \\
    & Previous Actions:id:0,type:DUAL\_POINT,text:G, ui\_type: ...& \textbf{id:1 view and click page content,state:finish} \\
    & Environment: id:0, text:Bass Headsets, type:TEXT ... & \textbf{id:2 type 'best rated headphones',state:unfinish} \\
    & Instruction: Search for the best rated headphones on Amazon. & \textbf{id:3 view and click page content,state:unfinish}\\
    & Answer: & \textbf{id:4 task complete,state:unfinish}\\
    &   & action: TYPE \\
    &   & text: best rated headphones \\
    
    \midrule
    {} & Given a mobile screen and a question, provide the action based on the screen information. & {} \\
    Llama 2& \textbf{SOP:} & \\
    & \hspace{0.0cm}\textbf{id:0 search on the website,state:ﬁnish} & \\
    +SOP& \hspace{0.0cm}\textbf{id:1 view and click page content,state:ﬁnish} & action: TYPE \\
    & \hspace{0.0cm}\textbf{id:2 type ’best rated headphones’,state:unﬁnish} & text: best rated headphones\\
    & \hspace{0.0cm}\textbf{id:3 view and click page content,state:unﬁnish} & \\
    & \hspace{0.0cm}\textbf{id:4 task complete,state:unﬁnish} & \\
    & Previous Actions:id:0,type:DUAL\_POINT,text:G, ui\_type: ... & \\
    & Environment: id:0, text:Bass Headsets, type:TEXT ...& \\
    & Instruction: Search for the best rated headphones on Amazon. & \\
    & Answer: & \\
    \bottomrule
    \end{tabular} 
        \caption{The model and corresponding prompts and responses. The first line details the construction of the Llama 2 sample, with bold text in other models highlighting the differences compared to Llama 2.}
        \label{prompt_statics}
    \end{table*}

\paragraph{SOP Details}
Even with the same instructions, the SOP for different tasks vary and are intricately linked to the specific execution operations of the current task. To distill key subtask sequences into standardized processes, we identified commonly executed actions and designated them as specific tasks. We excluded certain negative actions, such as 'going back' or 'closing,' and consolidated similar subtasks. Subsequently, in the current execution environment, each task was assigned a completion state.\\
For instance, If the semantic understanding of mobile screens \cite{sunkara2022towards} interprets the text context as the digit '9' or the letter 'g.', but the actual user action involves clicking on the Google icon, this task would be classified as 'search on the website.' For an in-depth definition of click types and their correlation with SOPs, please refer to the Appendix Table \Ref{sop_type}.

\paragraph{Implementation Details}
We conducted four sets of experiments: Llama 2, Llama 2+plan, Llama 2+plan+state, and Llama 2+SOP. The relationship among these experiments can be understood as follows: 'plan' requires the model to predict the overall plan during inference, 'plan+state' adds the completion status of each subtask to the 'plan', and 'SOP' can be viewed as transferring 'plan+state' to the input side, thereby guiding the model in generating executable instructions.\\
For each experiment, we utilized the computational power of 8 NVIDIA A100 GPUs during the training phase. The training process spanned 5 epochs with a learning rate set at 5e-5. Regarding the Google APP dataset, which is notably extensive, we sampled 10\% of the data. For other datasets, we conducted comprehensive training using the entire dataset.

\paragraph{Results}
The results of the comparative experiments are presented in Table \ref{aitw_result}. Our model exhibits the highest performance in all experiments, with its overall average metrics surpassing the previously best-performing model by 66.92\%, a significant increase of 1.49\%. Specifically, in individual domains, our model achieved superior results in four areas, while in the SINGLE domain, its performance is comparable to that of ChatGPT-4V.\\
From the perspective of the training process, models subjected to fine-tuning exhibit better performance compared to those relying solely on in-context predictions. This is because the fine-tuning process introduces more comprehensive and precise data into the model, as well as aligning LLM with our task instructions. In the in-context method, the multimodal ChatGPT-4V significantly surpasses pure LLM. This enhancement is attributed not only to the advanced capabilities of the model itself but also to the inclusion of direct image data, enriching the information fed into the model.\\
In experiments involving pure LLMs, the integration of a plan led to a noticeable decline in performance. Our analysis indicates that, on average, each instruction in the dataset is associated with 29 different operation pipelines, presenting a considerable challenge for the model to learn and accurately generate unique pipelines for each instruction. The complexity and variability of the data hinder the model’s ability to precisely fit the results. While adding the state status to each task atop the plan does introduce execution-relevant plan information into the model, the improvement observed is relatively minor.\\
We explored the factors contributing to the enhanced performance of models following the inclusion of SOP. First, SOP encapsulate historical operational behaviors and outline subsequent actions, enabling the model to more effectively comprehend the overall operational pipeline and gain a deeper insight into complex historical operations. Second, the introduction of SOP during the in-context stage significantly bolsters the model's inferential and predictive capabilities.

\begin{table*}[h]
    \centering
    \begin{tabular}{lllllll}
        \toprule
        Model &	Overall	& General &	Install	& GoogleApps &	Single &	WebShopping \\
        \midrule
      ChatGPT-CoT (5-shot) &	7.72 &	5.93 &	4.38 & 10.47 &	9.39 &	8.42 \\
      Palm 2	& 39.6	& -	& -	& -	& -	& - \\
      \midrule
      GPT-4V ZS +text	& 50.54	& 41.66	& 42.64	& 49.82	& 72.83	& 45.73 \\
      GPT-4V ZS image-only	& 51.92	& 42.44	& 49.18	& 48.26	& 76.34	& 43.35 \\
      GPT-4V ZS +history	& 52.96	& 43.01	& 46.14	& 49.18	& \textbf{78.29}	& 48.18 \\
      \midrule
      Llama 2	& 65.43	& 55.3	& 73.65	& 62.33	& 74.82	& 61.07 \\
      Llama 2+plan	& 62.08	& 52.1	& 71.65	& 56.23	& 74.18	& 56.22 \\
      Llama 2+plan+state	& 62.86	& 53.77	& 69.1	& 61.19	& 73.51	& 56.74 \\
      \midrule
      Llama 2+SOP	& \textbf{66.92}	& \textbf{55.8}	& \textbf{74.98}	& \textbf{63.95}	& 76.27	& \textbf{63.61} \\
      \bottomrule
    \end{tabular}
    \caption{Different Types of Instructions and Their Cor-
    responding Models’ Predictive Performance (task complete
    score).The Overall dataset calculates the average metrics of the five subsets.}
    \label{aitw_result}
\end{table*}
\paragraph{Computation Cost}
Table \ref{cost_time} compares the inference speed of the four models. 
Each model employs the vLLM framework \cite{kwon2023efficient} for accelerated predictions, with inference parameters configured to a maximum output length of 128 and a precision setting of float16. According to the comparison results, our model aligns with the original Llama 2 method in terms of speed and is ten times faster than the Llama 2+plan+state method.

\begin{table*}[h]
    \centering
\begin{tabular}{lll}
    \toprule
    Model &Inference (s/n)  & Token Length \\
    \midrule
    Llama 2 &  0.071 & 21 \\
    Llama 2+plan& 0.117 & 108  \\
    Llama 2+plan+state & 0.123 & 123 \\
    Llama 2+SOP & 0.071 & 21 \\
       
        \bottomrule
    \end{tabular}
    \caption{
        Computation Cost of Each Model: The computational efficiency of each model is determined by dividing the time taken (in seconds) by the number of inferences (n). The calculation of token length specifically relates to the response component.
    }
    \label{cost_time}
\end{table*}

\section{Conclusion}


In our automation execution application, we have integrated user interaction features to effectively manage privacy authorizations and related information. This is further enhanced by incorporating the flow of tasks as per SOP within the in-context. For offline training, we utilize a combination of samples, ensuring that this approach incurs no additional costs in practical usage. This method not only elevates the predictive performance baseline of the existing AitW dataset to a new level but also has potential applicability to similar models with operational workflows. To achieve heightened performance in this domain, directly inputting image data into multimodal large models is a key strategy. In the realm of multimodal systems, our future endeavor involves leveraging SOP to assist the model in comprehending historical operation sequences represented in image format.


\bibliographystyle{unsrt}
\bibliography{references}

\section{Appendix}

\paragraph{Sample Handling}

For the commands PRESS\_BACK, PRESS\_HOME, PRESS\_ENTER, STATUS\_TASK\_IMPOSSIBLE, STATUS\_TASK\_COMPLETE, and TYPE, our model's target output omits coordinate numbers and predicts only the command type. Additionally, for the TYPE command, input content is predicted. In double-point actions, if it's a click, the corresponding page element coordinates are mapped in the target output. For scroll actions, the scroll direction is determined from the coordinates and included in the target output.

\begin{table*}[h]
  \centering
\begin{tabular}{l|l|l}
  \toprule
  Action Type & Gesture Coordinate & Target Output \\
  \midrule
  PRESS\_BACK  & touch\_point:[-1.0, -1.0] lift\_point[-1.0, -1.0] & action type: PRESS\_BACK \\
  \midrule
  PRESS\_ENTER  & touch\_point:[-1.0, -1.0] lift\_point[-1.0, -1.0] & action type: PRESS\_ENTER \\
  \midrule
  PRESS\_HOME  & touch\_point:[-1.0, -1.0] lift\_point[-1.0, -1.0] & action type: PRESS\_HOME \\
  \midrule
  STATUS\_TASK\_COMPLETE  & touch\_point:[-1.0, -1.0] lift\_point[-1.0, -1.0] & action type: TASK\_COMPLETE \\
  \midrule
  STATUS\_TASK\_IMPOSSIBLE & touch\_point:[-1.0, -1.0] lift\_point[-1.0, -1.0] & action type: TASK\_IMPOSSIBLE \\
  \midrule
  TYPE & touch\_point:[-1.0, -1.0] lift\_point[-1.0, -1.0] & action type: TYPE \\ & typed\_text:XXX & text: XXX \\
  \midrule
  DUAL\_POINT (click) & touch\_point:[0.8497, 0.5964] lift\_point[0.8497, 0.5964] & action type: DUAL\_POINT \\ & & text: XXX type: ICON\_STAR id:1 \\
  \midrule
  DUAL\_POINT (scroll) & touch\_point:[0.8497, 0.5964] lift\_point[0.8497, 0.8964] & action type: SCROLL DOWN \\
   \bottomrule
  \end{tabular}
  \caption{the Enlarge-Element Method example: Transforming Action Coordinates into Page Elements.}
  \label{target_output}
\end{table*}
\paragraph{SOP Details}
This appendix, Table \ref{sop_type}, presents a detailed table outlining the SOP subtask that we have defined for all operations within the AitW dataset.

\begin{longtable}{l|l|l}
  \toprule
  Action & Text/ICON Type & SOP Task Description \\
  \midrule
  scroll  & - & scroll and view page content \\
  \midrule
  type  & * & type "*" \\
  \midrule
  status\_task\_complete  & - & task complete  \\
  \midrule
  press\_home  & - &  switch to the home screen \\
  \midrule
   & search & search on the website \\
   & search or type web address & search or type web address \\
   & g & search on the website \\
   & 9 & search on the website \\
   & add to cart & add goods to cart \\
   & accept \& continue & continue the next operate \\
   & agree & continue the next operate \\
   & ok & continue the next operate \\
   & remove & delete text \\
   & install & install the app \\
   & open & open the app \\
   & uninstall & uninstall the app \\
  	&	location	&	set your location		\\
	&	            search amazon	&	 search on the website		\\
dual\_point:TEXT	&	 search for anything	&	 search on the website		\\
	&	            search here 	&	 search on the website		\\
	&	            checkout	&	 checkout		\\
	&	            view in cart	&	 view in cart		\\
	&	            history	&	 view the history info		\\
	&	            add	&	 add goods to cart		\\
	&	            videos	&	open then video		\\
	&	            chrome	&	open the browser		\\
	&	            settings	&	display menu options		\\
	&	            site settings	&	display menu options		\\
	&	            m	&	display menu options		\\
	&	            notifications	&	check messages		\\
	&	            take me to gmail	&	use the email service		\\
\midrule
	&	            icon\_shopping\_cart		&	 view in cart	\\
	&	            icon\_plus		&	 add goods to cart	\\
	&	            icon\_three\_bars		&	 display menu options	\\
	&	            icon\_three\_dots		&	 display menu options	\\
	&	            icon\_v\_downward		&	 view page content	\\
	&	            icon\_mic		&	 input voice	\\
	&	            icon\_assistant		&	 open assistant	\\
	&	            icon\_play		&	 playing media	\\
	&	            icon\_person		&	 open profile and community	\\
dual\_point:ICON	&	            icon\_magnifying\_glass		&	 search on the website	\\
	&	            icon\_google		&	 search on the website	\\
	&	            icon\_chat		&	 send a message to someone	\\
	&	            icon\_settings		&	 set function	\\
	&	            icon\_nav\_bar\_rect		&	 switch to the other app	\\
	&	            icon\_nav\_bar\_circle		&	 switch to the home screen	\\
	&	            icon\_home		&	 switch to the home screen	\\
	&	            icon\_take\_photo		&	 take a photo	\\
	&	            icon\_time		&	 view time	\\
	&	            icon\_envelope		&	 view order	\\
	&	            icon\_location		&	set your location	\\
  \bottomrule
\caption{Types of actions in the AitW dataset and the corresponding SOP task descriptions designed for them.}
\label{sop_type}
\end{longtable}

\begin{figure}[]
  \centering
  \includegraphics[width=\textwidth]{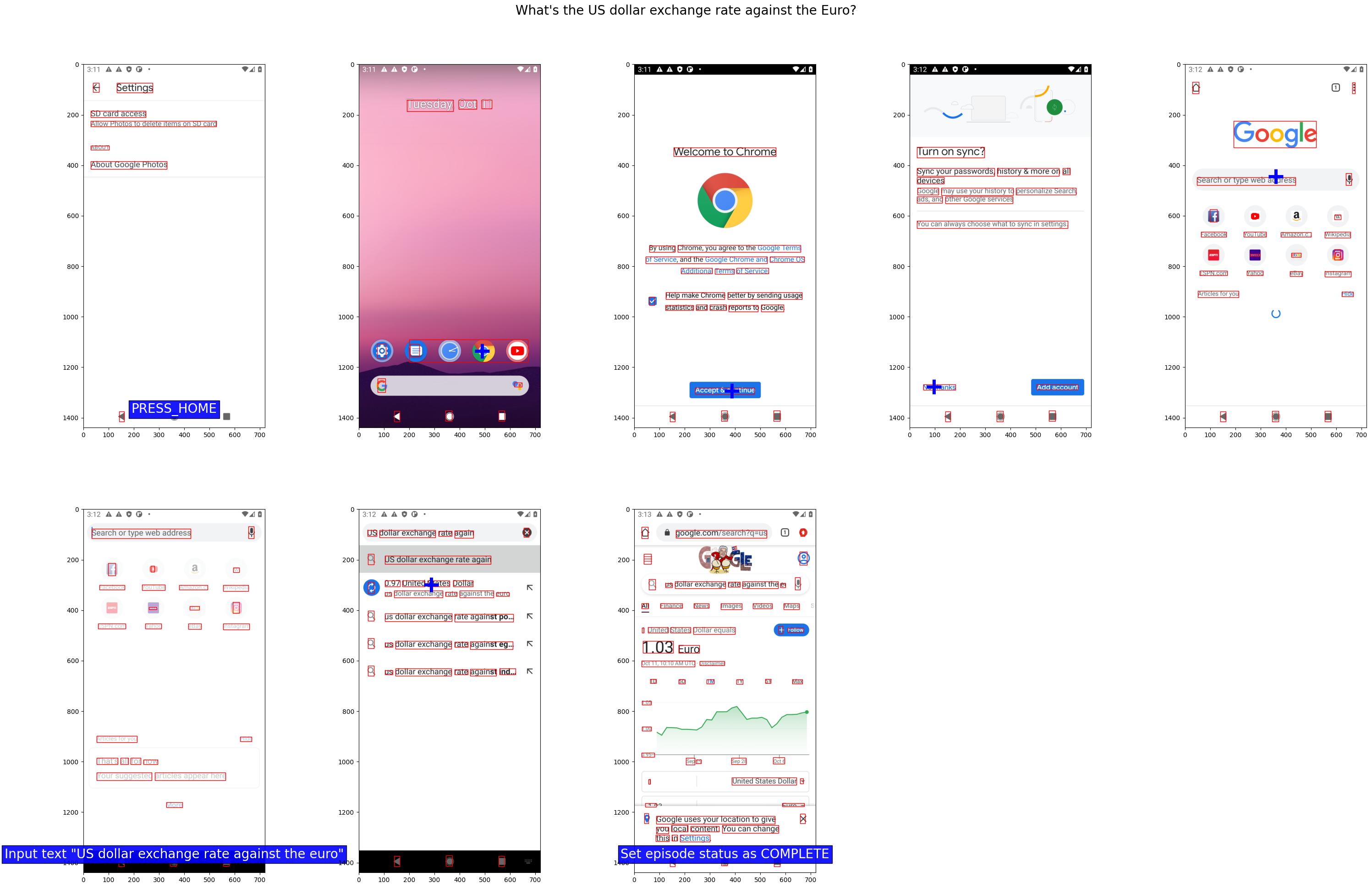} 
  \caption{An example episode from General.}
  \label{example_aitw}
\end{figure}

\paragraph{AitW Task}
We present task examples from the AITW benchmark dataset\cite{rawles2023android}, with Figure \ref{example_aitw} showcasing the 'General' subset, including tasks like 'What's the US dollar sxchange rate against the Euro?'.
\paragraph{Instruction Example}
Our designed model generates Instructions, which include Operational Instructions, Task Status Instructions, and Structured Instructions.
  \begin{figure}[]
    \centering
    \includegraphics[width=0.78\textwidth,keepaspectratio]{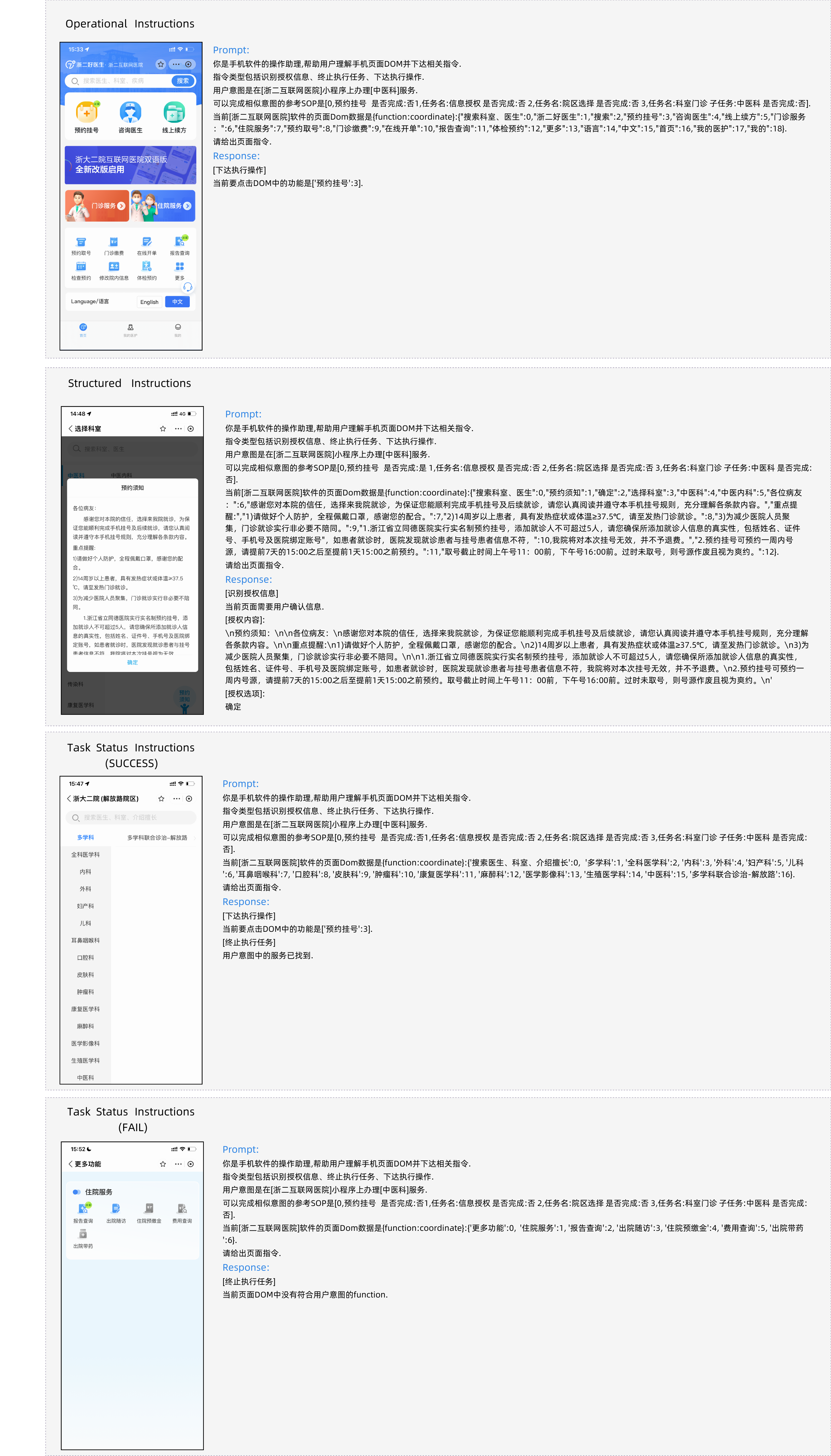} 
    \caption{The instrucion example.
    }
    \label{instructions}
\end{figure}
\end{document}